\documentclass[onecolumn,showpacs,preprintnumbers,amsmath,amssymb]{revtex4}

\usepackage{graphicx}
\usepackage{amsmath}
\usepackage{bm}
\usepackage{multirow}
\usepackage{comment}

\newcommand{\simgt}{\lower.5ex\hbox{$\; \buildrel > \over \sim \;$}}
\newcommand{\simlt}{\lower.5ex\hbox{$\; \buildrel < \over \sim \;$}}
\newcommand{\largesmall}{\lower.5ex\hbox{$\; \buildrel {\LARGE >} \over {\small 
<}$}}

\def\geffxi{{\xi}}

\begin{document}
\title{
Characterizing the linear growth rate of cosmological 
density perturbations in an $f(R)$ model}

\author{
Tatsuya {Narikawa} and Kazuhiro {Yamamoto}}

\affiliation{
Department of Physical Science, Hiroshima University,
Higashi-Hiroshima 739-8526,~Japan}


\begin{abstract}
We investigate the linear growth rate of cosmological matter density
perturbations in a viable $f(R)$ model both numerically and 
analytically. We find that the growth rate in the scalar-tensor
regime can be characterized by a simple analytic formula. 
We also investigate a prospect of constraining the Compton
wavelength scale of the $f(R)$ model with a future weak 
lensing survey. 
\end{abstract}

\preprint{HUPD-0907}
\maketitle

\section{Introduction}

Cosmological observations of distant Ia supernovae discovered
that our Universe is undergoing an accelerated expansion period 
\cite{Riess98,Perlmutter99}, which is supported by other
observations of the cosmic microwave background anisotropies 
and the large scale structure of galaxies 
\cite{WMAP3,Einsenstein05,Percival07}. 
These observations are explained by the cosmological model 
with the cosmological constant $\Lambda$. The cosmological 
constant can be regarded as the vacuum energy, however,
the smallness of the observed value raises a fine-tuning problem
\cite{Weinberg}.
To explain the cosmic accelerated expansion, many dark energy
models have been proposed 
(see e.g., \cite{Peebles,DE06} and the references therein). 
Modification of the gravity theory is an alternative approach, 
for example, 
$f(R)$ model \cite{Carroll,Nojiri,Capozziello,Sotiriou08}
and the Dvali, Gabadadze, and Porrati (DGP) model in the
context of the braneworld scenario \cite{DGP}.

Many authors have studied dynamical dark energy models
\cite{Huterer01,Chevallier,Linder03,Sahni08}. 
Dynamical dark energy models may have similar expansion rates to models 
of modified gravity, because modification of the gravity theory may 
affect the background expansion history. 
Therefore, the observations of the background expansion history alone 
are unable to distinguish between modified gravity and dynamical 
dark energy.
The key to distinguish between modified gravity and dynamical dark energy
is the growth of cosmological perturbations \cite{Ishak06,Yamamoto07,
Huterer07,Kunz07,Song09,Koyama09}.
The growth history of cosmological perturbations can 
be tested with the large scale structure in the Universe.
Many projects of large survey of galaxies are in progress 
or planned \cite{DETF06,Frieman08,sdss3,sumire,lsst,ska,
Robberto,DES,Pan-STARRS,Euclid}, and
these surveys might give us a hint in exploring the origin of 
the accelerated expansion of the Universe and the nature of 
gravity \cite{LinderRD,Nesseris08,Porto08,Heavens07,Zhang07,Amendola08,
Schmidt08,Song08,Tsujikawa08a,Thomas09,Zhao09a,Zhao09b,
Guzik09,Bean09}.

Cosmological perturbations in modified gravity 
models have been investigated by many authors \cite{
Linder05,Knox06,Linder07,Gong08,Sereno06,Wang08,Polarski08, 
Ballesteros08,Gannouji08a,Gannouji08b,Bertschinger08,
Laszlo08,
Wei08,Dent09a,Dent09b}. 
In the present paper, we investigate the growth history of 
matter density perturbations in $f(R)$ models. 
$f(R)$ model is a modified gravity model, constructed 
by replacing the gravitational Lagrangian with a general 
function of the Ricci scalar $R$.
The viable $f(R)$ models have been proposed 
\cite{Starobinsky07,HuSawicki07,Battye,Tsujikawa08,Tsujikawa09}, 
which explain the late-time accelerate expansion of the background 
Universe, and satisfy the local gravity constraints.
The viable model which also explains an inflationary epoch 
in the early Universe is extensively proposed
\cite{Nojiri07a,Nojiri07b,Cognola}.  
For the local gravity constraints, the chameleon mechanism 
is supposed to play an important role 
\cite{Mota04,Khoury04,Brax08}.  
By this mechanism, a field that modifies the gravity is hidden 
in the local region with high density. 
We note that a problem of the theory in the strong gravity regime
is under debate \cite{Kobayashi,Upadye}. 
Though the evolution of cosmological perturbations in $f(R)$ models 
has been studied so far \cite{Pogosian,Song07,Bean07,Tsujikawa07,delaCD,Motohashi,Brax09,Zhang06,Capozziello09,Carloni09}, 
our investigation is focused on a new description of the growth rate 
for the $f(R)$ model.

This paper is organized as follows. In Sec.~I\hspace{-.1em}I, we briefly 
review the viable $f(R)$ models.
In Sec.~I\hspace{-.1em}I\hspace{-.1em}I, we investigate the evolution of density perturbations 
in the $f(R)$ model both numerically and analytically. 
We find that the growth rate of density perturbations 
can be characterized by a simple analytic formula, 
which approximately describes the growth rate in the scalar-tensor regime. 
The growth rate in the general-relativity regime 
is also investigated. 
In Sec.~I\hspace{-.1em}V, we investigate a future prospect of constraining 
the $f(R)$ model assuming a future large survey of 
weak lensing statistics on the basis of the Fisher matrix analysis.
Section V is devoted to summary and conclusions.
Throughout the paper, we use the unit in which the 
speed of light equals 1 and $\hbar=1$. 

\section{a brief review of $f(R)$ model}
We briefly review $f(R)$ model, which is defined by the action,
\begin{equation}
 S=\frac{1}{16\pi G}\int d^4x\sqrt{-g}(R+f(R))
+\int d^4x\sqrt{-g}L^{(\rm m)},
\label{action}
\end{equation}
where $G$ is the gravitational constant, 
and $L^{(\rm m)}$ is the matter Lagrangian density. 
We consider the viable models, proposed in Refs.~
\cite{Starobinsky07,HuSawicki07,Battye,
Tsujikawa08,Tsujikawa09}.
The viable models have an 
asymptotic formula at the late-time Universe {($R\gg R_c$)}, 
which can be written as 
\begin{equation}
  f(R)=-\lambda R_c\left[1-\left({R_c\over R}\right)^{2n}\right],
\label{fR}
\end{equation}
where $R_c$ is a positive constant whose value is the same 
order as that of the present Ricci scalar, 
and $\lambda$ is a nondimensional constant. 
Because the term $\lambda R_c$ plays a role of the cosmological 
constant, we may write $\lambda R_c=6(1-\Omega_0)H_0^2$, 
where $H_0$ is the Hubble constant and $\Omega_0$ is the matter 
density parameter. 
Note that we assume the spatially flat Universe. 

It is well known that $f_R=d f(R)/d R$ plays 
a roll of a new degree of freedom, which behaves like a 
scalar field with the mass 
\begin{eqnarray}
  m^2={1\over 3} \left({1+f_R\over f_{RR}}-R\right),
\label{m2}
\end{eqnarray}
where we defined $f_{RR}={d^2 f(R)/dR^2}$.
Assuming $|f_R|\ll 1$ and $R f_{RR}\ll 1$ for the viable model, 
the mass is simply $m^2=1/(3f_{RR})$.

We focus on the evolution of matter density perturbations 
in the $f(R)$ model, whose Fourier coefficients obey
(e.g., \cite{Tsujikawa09})
\begin{equation}
 \ddot \delta +2H \dot \delta -4\pi G_{\rm eff} \rho \delta=0,
\label{evolutioneq}
\end{equation}
where the dot denotes the differentiation with respect to the 
cosmic time, $H=\dot a/a$ is the Hubble parameter, $\rho$
is the matter mean density, and $G_{\rm eff}$ is the 
effective gravitational constant, which is written as
\begin{equation}
 {G_{\rm eff}\over G}=1+{1\over3}{k^2/a^2\over k^2/a^2+1/(3f_{RR})},
\label{GeffG}
\end{equation}
where $k$ is the wave number, and $a$ is the scale factor 
normalized to unity at present epoch (cf. \cite{Zhang06}).
As is noted in the above, the physical meaning of $m^2=1/(3f_{RR})$ 
is the square of the mass of the new degree of freedom which 
modifies the gravity force. We have the general-relativity 
regime, $G_{\rm eff}=G$,  for $k/a\ll m$, and the scalar-tensor
regime, $G_{\rm eff}=4G/3$, for  $k/a\gg m$, respectively.
Thus, the evolution of matter density perturbations 
depends on the wavenumber $k$, whose behavior is determined 
by the mass $m^2=1/(3f_{RR})$. 

For the Einstein de Sitter universe, the exact solution of 
Eq.~(\ref{evolutioneq}) is found in the literature \cite{Motohashi}. 
However, we consider the low density universe, where the solution 
of Eq.~(\ref{evolutioneq}) is described in a different form 
in comparison with that of \cite{Motohashi}. 
From Eq.~(\ref{fR}), we have
\begin{eqnarray}
&&  f_{RR}={d^2f(R)\over dR^2}=2n(2n+1)\lambda R_c{R_c^{2n}\over R^{2n+2}}.
\end{eqnarray}
Furthermore, using the formulas $\lambda R_c=6(1-\Omega_0)H_0^2$ and
$R=3H_0^2\left[{\Omega_0/a^3}+4(1-\Omega_0)\right]$, we have
\begin{equation}
{1\over3f_{RR}}={\Omega_0H_0^2\over 4n(2n+1)}
\left({\lambda\over2}\right)^{2n} \left({\Omega_0\over1-\Omega_0}\right)^{2n+1}
\left({1\over a^3}+{4(1-\Omega_0)\over 
\Omega_0}\right)^{2n+2}.
\label{abc}
\end{equation}
Denoting the wavenumber corresponding to the Compton wavelength
$1/m$ at the present epoch by $k_C$, 
\begin{equation}
k_C^2={\Omega_0H_0^2\over 4n(2n+1)}
\left({\lambda\over2}\right)^{2n} \left({\Omega_0\over1-\Omega_0}\right)^{2n+1}
\left({1}+{4(1-\Omega_0)\over 
\Omega_0}\right)^{2n+2},
\label{defkc}
\end{equation}
Equation (\ref{abc}) is rewritten as
\begin{equation}
 {1\over3f_{RR}}=k_C^2 
\left({\Omega_0{a^{-3}}+{4(1-\Omega_0)}\over {\Omega_0}
+{4(1-\Omega_0)}}\right)^{2n+2}.
\label{kcintro}
\end{equation}


We denote the growth factor by $D_1(a,k)$, which is the solution of 
Eq.~(\ref{evolutioneq}) normalized so as to be 
$D_1(a,k)\simeq a$ at $a\ll1$. The growth rate is defined  by 
\begin{equation}
f(a,k)={d\log D_1(a,k)\over d\log a}.
\end{equation}
Using the growth rate $f(a,k)$, Eq.~(\ref{evolutioneq}) is rephrased as
\begin{equation}
{d f\over d\ln a}
+f^2+\left(2+{\dot H\over H^2}\right)f={3\over 2}{G_{\rm eff}\over G}\Omega_m(a),
\label{evolutionf}
\end{equation}
where 
$\Omega_m(a)$ is defined by 
$\Omega_m(a)={H_0^2 \Omega_0 a^{-3}/ H^2}$.
We assume that the background expansion is well approximated by 
the $\Lambda$CDM model, where the Hubble parameter satisfies
\begin{equation}
{\dot H\over H^2}=-{3\over2}\Omega_m(a),
\label{eqH}
\end{equation}
and the energy conservation equation 
\begin{equation}
{d\Omega_m(a)\over d\ln a}=-3\Omega_m(a)(1-\Omega_m(a)).
\label{eneconeq}
\end{equation}
Using Eqs.~(\ref{evolutionf})-(\ref{eneconeq}) yields
\begin{equation}
-3\Omega_m(a)(1-\Omega_m(a)){d f\over d\Omega_m(a)}
+f^2+\left(2-{3\over 2}\Omega_m(a)\right)f={3\over 2}{G_{\rm eff}\over G}\Omega_m(a),
\label{evolutionfeq}
\end{equation}
which is useful to find an approximate
solution, as we see in the next section. 

\section{growth of density perturbations in $f(R)$ model}
In this section, we investigate the evolution of matter density perturbations 
in the $f(R)$ model. 
In Sec.~I\hspace{-.1em}I\hspace{-.1em}I A, we consider 
the scalar-tensor regime, $k/a\gg m$, in which the wavelength is shorter than
the Compton wavelength. 
In Sec.~I\hspace{-.1em}I\hspace{-.1em}I B, we consider the
general-relativity regime, $k/a\ll m$, in which the wavelength is larger than
the Compton wavelength. 

\subsection{scalar-tensor regime}
In the scalar-tensor regime, $k/a\gg m$,  the effective gravitational 
constant becomes $G_{\rm eff}=4G/3$. In this case, we find that 
Eq.~(\ref{evolutionfeq}) has the solution expressed in the form 
\cite{footnote1} 
\begin{eqnarray}
f(a,k)=f_0 \Omega_m(a)^{\widetilde\gamma(a)},
\label{STregime}
\end{eqnarray}
where $f_0$ obeys $f_0^2+f_0/2=2$, therefore $f_0=(-1+\sqrt{33})/4$,
and 
\begin{eqnarray}
\widetilde\gamma(a)=\sum_{\ell=0}\zeta_\ell (1-\Omega_m(a))^\ell,
\end{eqnarray}
where $\zeta_\ell$ is the expansion coefficients. 
The first few terms of $\widetilde\gamma(a)$ are
\begin{equation}
\widetilde\gamma(a)= {9-\sqrt{33}\over6}
 -{93-17\sqrt{33}\over666}\left(1-\Omega_m(a)\right) 
  +{\cal O}\left((1-\Omega_m(a))^2\right).
\end{equation}
This can be generalized to the case when $G_{\rm eff}/G(=\geffxi)$ 
is a constant value, in which the solution of Eq.~(\ref{evolutionf}) 
has the same formula as that of (\ref{STregime}) but with 
$
f_0 = ({-1+\sqrt{1+24\geffxi}})/{4}
$
and
\begin{eqnarray}
 &&{\widetilde\gamma}(a)=\frac{-41+24\geffxi+\sqrt{1+24\geffxi}}
{-70+48\geffxi} 
+{1\over 8(-143+24\geffxi)(-35+24\geffxi)^2}
\nonumber
\\
  &&~~~~~~
  \times\Bigl[(-41+24\geffxi+\sqrt{1+24\geffxi})\bigl(
347-17\sqrt{1+24\geffxi}
+24\geffxi(-13+\sqrt{1+24\geffxi})\bigr)\Bigr] \nonumber \\
 &&~~~~~~
  \times(1-\Omega_m(a))+{\cal O}\left((1-\Omega_m(a))^2\right).
\end{eqnarray}
Here we assume that $G_{\rm eff}/G(=\geffxi)$ is constant,
but we utilize this formula by replacing $\geffxi$
with the right-hand-side of Eq.~(\ref{GeffG}).

\begin{figure}[htbp]
\begin{center}
\includegraphics[width=15cm,height=7.5cm,clip]{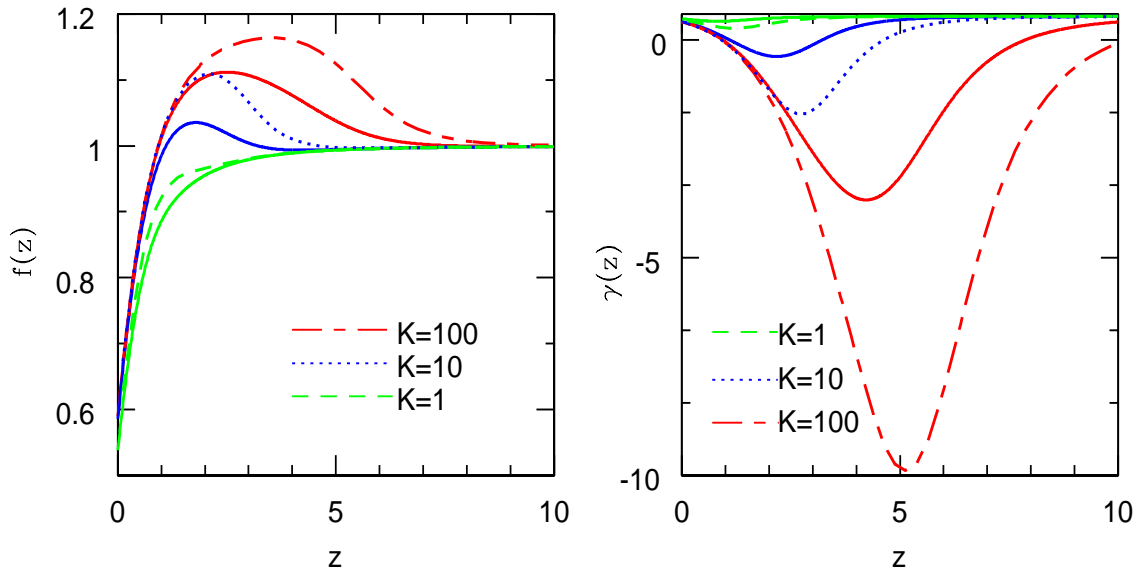}
\caption{Left panel: $f(a,k)$ as a function of $z(=1/a-1)$ for the model $n=1$. 
The solid curves are obtained by solving the differential 
equation (\ref{evolutionf}) numerically, for $K(=k/k_C)=10^2$, $10$, 
and $1$, respectively, from the top to the bottom. 
The long dashed curve, the dotted curve and the short dashed curve adopt 
the approximate formula, for $K(=k/k_C)=10^2$, $10$, and $1$, respectively, 
from the top to the bottom. 
Right panel: The growth index $\gamma(a,k)$ as a function of $z(=1/a-1)$, 
corresponding to the left panel. The parameter of the curves is 
the same as that of the left panel. The curves correspond to
$K(=k/k_C)=10^2$ and $10$, $1$ respectively, from the bottom to the
top.
}
\label{fig:f.gamma}
\end{center}
\end{figure}
\begin{figure}[htbp]
\begin{center}
\includegraphics[width=15cm,height=7.5cm,clip]{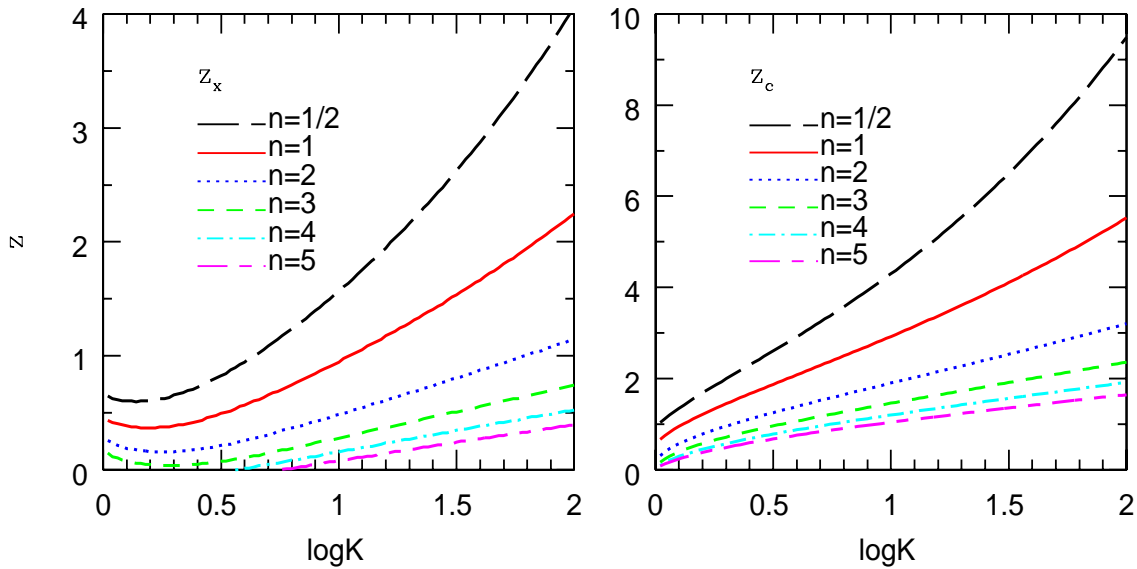}
\caption{Left panel: Redshift $z_x$ when the difference of 
the growth rate becomes $f^{(\rm appr)}-f^{(\rm exac)}=0.03$, 
as a function of $K(=k/k_C)$.
The curves are $n=1/2,~1,~2,~3,~4$, and $5$, respectively, 
from the top to the bottom. 
Right panel: Transition redshift $z_c$ as a function of $K(=k/k_C)$. 
From the top to the bottom, curves are $n=1/2,~1,~2,~3,~4$, and 
$5$, respectively. 
Here we adopted the background expansion of the Universe 
is the $\Lambda$CDM model with $\Omega_0=0.28$.
}
\label{fig:stregime}
\end{center}
\end{figure}

The left panel of Fig.~\ref{fig:f.gamma} shows
the growth rate $f(a,k)$ as a function of the redshift $z(=1/a-1)$.
Here we adopted $n=1$. The solid curves are obtained by solving 
Eq.~(\ref{evolutionf}) numerically, for  
$K(=k/k_C)=10^2,~10,~1$, from the top to the bottom, respectively. 
Here we assumed the background expansion of the Universe 
is the $\Lambda$CDM model with $\Omega_0=0.28$.
The dot-dashed curve, the dotted curve, and the short dashed curve are 
the approximate formula up to 1st order of $(1-\Omega_m(a))$, 
for the wavenumber $K(=k/k_C)=10^2,~10,~1$, respectively.
In the computation of the approximate formula, we adopted 
the right-hand-side of Eq.~(\ref{GeffG}) as $\xi$.  
One can see that the approximate solution approaches the 
exact solution at the late time of the redshift.


Following the previous works 
(e.g., see \cite{Polarski08}), the growth index 
$\gamma(a,k)$ is introduced by 
\begin{equation}
f(a,k)=\Omega_m(a)^{\gamma(a,k)},
\label{fgammadef}
\end{equation}
which is related with $\widetilde\gamma(a,k)$ by
\begin{equation}
{\gamma(a,k)={\ln f_0\over \ln\Omega_m(a)}+\widetilde\gamma(a,k)}.
\label{gammadef}
\end{equation}
The behavior of the growth index $\gamma(a,k)$ 
in the scalar-tensor regime 
is well approximated by Eq.~(\ref{gammadef}), as is 
demonstrated in the right panel of Fig.~\ref{fig:f.gamma}, 
which plots $\gamma(a,k)$ as a function of 
the redshift $z$, for the wavenumber $K(=k/k_C)=10^2,~10,~1$, 
from the bottom to the top, respectively. 
The solid curves are obtained by solving 
Eqs.~(\ref{evolutionf}) and (\ref{fgammadef}) numerically, for  
$K(=k/k_C)=10^2,~10,~1$, from the bottom to the top, respectively. 
The dot-dashed curve, the dotted curve, and the short dashed curve 
are the approximate solution of $\widetilde\gamma(a,k)$ and (\ref{gammadef}), 
for  $K(=k/k_C)=10^2,~10,~1$, respectively. 
One can see that the approximate solution approaches the 
exact solution at the late time of the redshift, however, 
the validity is limited to the late time of the small redshift.

Let us discuss the valid region of the approximate solution. 
The left panel of Fig.~\ref{fig:stregime} plots the redshift $z_x$
as a function of $K(=k/k_C)$ for $n=1/2,~1,~2,~3,~4$, and $5$, respectively,
from the top to the bottom, where $z_x$ is defined by the redshift 
when the difference of the growth rate becomes 
$f^{(\rm appr)}-f^{(\rm exac)}=0.03$. 
Here $f^{(\rm exac)}$ is the exact solution obtained by solving 
Eq.~(\ref{evolutionf}) numerically, while $f^{(\rm appr)}$ is 
the approximate solution. Thus, the approximate solution of
the growth rate approaches the exact solution after the redshift 
$z_x$, which depends on $k/k_C$ as well as $n$. 
As $n$ is larger or $k/k_C$ is smaller, $z_x$ becomes smaller. 
For the case $n\geq 4$ and smaller value of $k/k_C$, we have 
no solution of $z_x$.

The above behavior is related with the transition redshift $z_c$, 
when the scalar-tensor regime starts, which
we defined by $k(1+z_c)=m$, i.e., 
\begin{equation}
{k^2(1+z_c)^2}=k_C^2\left({{\Omega_0(1+z_c)^{3}}+{4(1-\Omega_0)
}\over {\Omega_0}+{4(1-\Omega_0)}}\right)^{2n+2}.
\end{equation}
%
The right panel of Fig.~\ref{fig:stregime} plots $z_c$ as function of $K(=k/k_C)$ 
for $n=1/2,~1,~2,~3,~4,~5$, respectively, from the top to the bottom. 
Figure \ref{fig:stregime} shows $z_x<z_c$. Thus 
the approximate formula approaches the exact solution after 
the scalar-tensor regime starts. 
For the model with larger value of $n$, the 
Compton scale evolves rapidly. Then, the transition redshift 
$z_c$ becomes small as $n$ becomes large.
For the smaller value of $K(=k/k_C)$, the transition redshift 
$z_c$ becomes smaller. 
This is the reason why $z_x$ is smaller, as 
$n$ is larger or $k/k_C$ is smaller. 
Therefore, for the case when $n$ is large and $k/k_C$ is smaller, 
the redshift when the approximate formula starts to work becomes 
later. For the case $n\simlt 2$,  the late-time behavior of the
growth rate can be approximated by the approximate formula as long as 
$K\simgt1$.

We here mention the relation between the parameter $k_C$ and 
the parameter $f_{R0}$ adopted in Refs.~
\cite{HuSawicki07,Schmidt09},  in which the case $n=1/2$ is investigated. 
In this case, $|f_{R0}|=2(1-3\Omega_0/4)H_0^2/k_C^2$. 
For $|f_{R0}|\simeq 10^{-4}-10^{-6}$, we have $k_C\simeq 0.04-0.4~h 
{\rm Mpc}^{-1}$. The scalar-tensor regime appears rather earlier in this model, 
as shown in Fig.~\ref{fig:stregime}.

\subsection{general-relativity regime}
In this subsection, we consider the growth rate of density perturbations
at the early time epoch of the Universe, $a\ll1$, adopting the approximation, 
\begin{equation}
{1\over3f_{RR}}={\Omega_0H_0^2\over 4n(2n+1)}
\left({\lambda\over2}\right)^{2n} \left({\Omega_0\over1-\Omega_0}\right)^{2n+1}
\left({1\over a}\right)^{6n+6},
\end{equation}
which yields the simple form of the effective gravitational constant
\begin{equation}
 {G_{\rm eff}\over G}=1+{1\over3}{k^2\over k^2+k_0^2 a^{-3N}},
\end{equation}
where
\begin{eqnarray}
&& N=2n+{4\over 3},
\\
&&
k_0^2={\Omega_0H_0^2\over 4n(2n+1)}
\left({\lambda\over2}\right)^{2n} \left({\Omega_0\over1-\Omega_0}\right)^{2n+1}.
\end{eqnarray}
As mentioned in the previous section, $3f_{RR}$ has the meaning 
of the square of the Compton wavelength. Thus this model can 
be regarded as the model that the Compton wavelength simply 
evolves as $1/m=a^{3N/2+1}/k_0$. 

In the case when $N$ is a positive integer, we derive an approximate
solution of Eq.~(\ref{evolutionf}) in an analytic manner.
With the use of (\ref{fgammadef}), one can rewrite 
Eq.~(\ref{evolutionfeq}) as
\begin{eqnarray}
 -3\Omega_m(a)(1-\Omega_m(a))\ln\Omega_m(a)\frac{d\gamma}{d\Omega_m(a)}
 +3\Omega_m(a)\left(\gamma-\frac{1}{2}\right)+\Omega_m(a)^\gamma
 -\frac{3}{2}\frac{G_{\rm eff}}{G}\Omega_m(a)^{1-\gamma}
  -3\gamma+2=0.
\label{req}
\end{eqnarray}
In a straightforward manner, we find the solution for 
$\gamma(a,k)$ expanded in terms of $1-\Omega_m(a)$, as follows.
\begin{eqnarray}
  \gamma(a,k)=\sum_{\ell=0} \zeta_\ell(a,k) (1-\Omega_m(a))^\ell,
\end{eqnarray}
where $\zeta_\ell(a,k)$ is the expansion coefficient. 
For example, for $N=1$, we find
\begin{eqnarray}
  &&\displaystyle{\gamma(a,k)={6\over 11} 
- \frac{K^2 \Omega_0}{11 (1- \Omega_0)} }
  \nonumber \\
 &&\hspace{1.2cm}\displaystyle{
 +\left[\frac{15}{2057} - \frac{131 K^2 \Omega_0}{ 4114 (1 - \Omega_0)} 
 + \frac{263 K^4 \Omega_0^2}{ 4114 (1 - \Omega_0)^2} \right]
(1-\Omega_m(a))+{\cal O}\left((1-\Omega_m(a))^2\right)}, 
\label{gammaN1}
\end{eqnarray}
where $K=k/k_0$. We also have 
\begin{eqnarray}
 &&\gamma(a,k)=\displaystyle{\frac{6}{11} + \left[\frac{ 15}{2057} 
 - \frac{K^2 \Omega_0^2}{ 17 (1 - \Omega_0)^2}\right](1-\Omega_m(a)) }
 ~~~~~~~~~~~~~~~~~~~~~~~~~~~~~~~~
 \nonumber \\
 && \hspace{1.2cm} \displaystyle{+\left[\frac{4205}{1040842} - \frac{643 K^2 \Omega_0^2}
 { 8602 (1 - \Omega_0)^2}\right](1-\Omega_m(a))^2
  +{\cal O}\left((1-\Omega_m(a))^3\right),} 
\label{gammaN2}
\end{eqnarray}
for $N=2$,
\begin{eqnarray}
 &&\gamma(a,k)=\displaystyle{\frac{6}{11}+\frac{15}{2057}(1-\Omega_m(a))}
 ~~~~~~~~~~~~~~~~~~~~~~~~~~~~~~~~~~~~~~~~~~~~~~~~~~~~~~~~~~~
 \nonumber \\
 && \hspace{1.2cm} \displaystyle{+\left[\frac{4205}{1040842}
-\frac{K^2 \Omega_0^3}
{23 (1 - \Omega_0)^3}\right](1-\Omega_m(a))^2
  +{\cal O}\left((1-\Omega_m(a))^3\right)}, ~~~
\label{gammaN3}
\end{eqnarray}
for $N=3$,
\begin{eqnarray}
 &&\gamma(a,k)=\displaystyle{\frac{6}{11}+\frac{15}{2057}(1-\Omega_m(a))
  +\frac{4205}{1040842}(1-\Omega_m(a))^2} 
 ~~~~~~~~~~~~~~~~~~~~~~~~~~~~~
 \nonumber \\
 && \hspace{1.2cm} \displaystyle{+ \left[\frac{31449595}{11288972332}-\frac{K^2\Omega_0^4}
 {29(1-\Omega_0)^4}\right](1-\Omega_m(a))^3
  +{\cal O}\left((1-\Omega_m(a))^4\right)}, 
\label{gammaN4}
\end{eqnarray}
for $N=4$, respectively.

\begin{figure}[htbp]
\begin{center}
\includegraphics[width=10cm,height=10cm,clip]{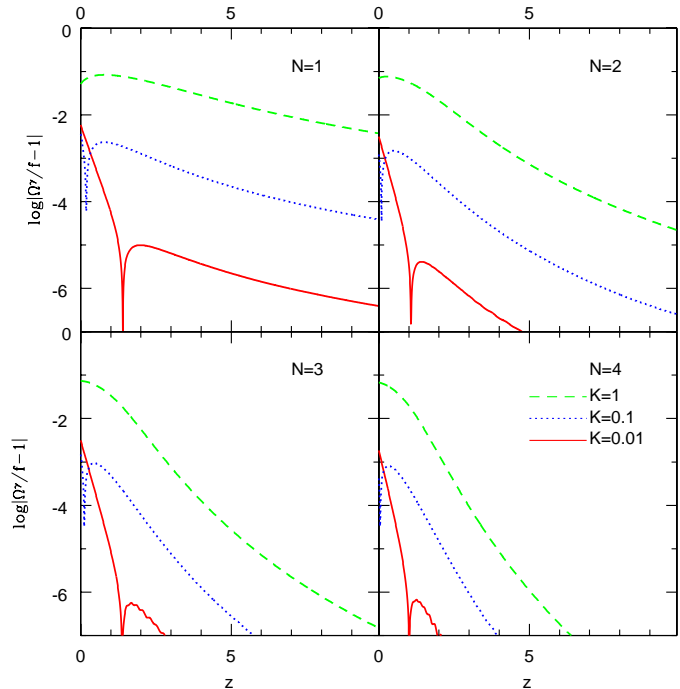}
\caption{The relative difference of the growth factor 
between the exact solution $f$, which is obtained by solving 
Eq.~(\ref{evolutionf}) numerically, and the approximate solution in the form 
$\Omega_m^\gamma$, as a function of $z$. The four panels correspond to
$N=1,~2,~3,$ and $4$, respectively.
In each panel, the curves correspond to $K=0.01$, $0.1$, $1$, respectively, 
from the bottom to the top. }
\label{fig:GRregime}
\end{center}
\end{figure}

Figure \ref{fig:GRregime} demonstrates the validity of the approximate 
formulas, by plotting the relative difference between the exact 
solution $f$ and the approximate solution $\Omega_m^\gamma$ as a function 
of the redshift. 
The four panels assume $N=1,~2,~3,~4$, respectively. In each panel,
the cases of the wavenumber $K(=k/k_C)=1,~0.1,~0.01$, are plotted. 
The solid curve, the dotted curve and the dashed curve correspond to 
$K=1,~0.1,~0.01$, respectively.
For $N=1$, we used the approximate formula (\ref{gammaN1}) up to the $1$st 
order of $(1-\Omega_m(a))$. 
For $N=2$, we used the approximate formula (\ref{gammaN2}) up to the $2$nd 
order of $(1-\Omega_m(a))$. 
For $N=3$, we used the approximate formula (\ref{gammaN3}) up to the $2$nd 
order of $(1-\Omega_m(a))$. 
For $N=4$, we used the approximate formula (\ref{gammaN4}) up to the $3$rd 
order of $(1-\Omega_m(a))$. 
Even if we adopted higher order term of (\ref{gammaN1})$\sim$(\ref{gammaN4}), 
the approximate formula only slightly improves the accuracy for $K=1$. 
The approximate formula is valid for $K\simlt 0.1$.

\section{Constraint on $f(R)$ model from weak lensing survey}
Cosmological constraints on the $f(R)$ model have been 
investigated in Refs.~\cite{HuSawicki07,Schmidt09,Girones,Yamamoto10}.
The weak lensing statistics is useful to obtain a constraint on the 
growth history of cosmological density perturbations 
observationally. 
We now consider a prospect of constraining the $f(R)$ model
with a future large survey of the weak lensing. 
To this end, we adopt the Fisher matrix analysis, which 
is frequently used for estimating minimal attainable
constraint on the model parameters. 
To be self-contained, we summarized 
the fisher matrix analysis in the Appendix 
(see also \cite{Yamamoto07}, and the references therein).
Here we focus on the constraint on the Compton wavenumber 
parameter $k_C$ defined by Eq.~(\ref{defkc}) or (\ref{kcintro}). 
In this analysis, we obtained the 
growth rate and the growth factor by numerically 
solving Eq.~(\ref{evolutionf}) and 
\begin{eqnarray}
 D_1(a,k)=a\exp\left[\int_0^a{da'\over a'}(f(a',k)-1)\right],
\end{eqnarray}
without using the approximate formula.

We briefly review how the signal of the weak lensing reflects 
the modification of the gravity in the $f(R)$ model. 
In the Newtonian gauge, the metric perturbations of the Universe 
can be describe by the curvature perturbation $\Phi$
and the potential perturbation $\Psi$, 
\begin{eqnarray}
ds^2=-\left(1+2\Psi\right)dt^2+a^2(t)\left(1+2\Phi\right)d{\bf x}^2.
\end{eqnarray}
In the $f(R)$ model, the relations between the two metric potentials 
and the matter density perturbations are altered. 
In the subhorizon limit, the $f(R)$ model yields (e.g., 
\cite{Silvestri} and references therein)
\begin{eqnarray}
k^2\Psi&\equiv&-4\pi G\mu(a,k) a^2\rho
\delta\label{Poissoneq} \\
{\Phi\over\Psi}&\equiv&-\nu(a,k)
\label{gslip},
\end{eqnarray}
with
\begin{eqnarray}
\mu(a,k)&=&{a^2+4f_{RR}k^2\over a^2+3f_{RR}k^2}, \\
\nu(a,k)&=&{a^2+2f_{RR}k^2\over a^2+4f_{RR}k^2},
\end{eqnarray}
where we used $|f_R|\ll 1$ and $R f_{RR}\ll 1$. 
Equation (\ref{Poissoneq}) is the modified Poisson equation. 
In general relativity, $\mu=\nu=1$.
%
With Eqs. (\ref{Poissoneq}) and (\ref{gslip}), we have
\begin{equation}
  k^2(\Phi-\Psi)=8\pi Ga^2\rho \delta.
\label{PPd}
\end{equation}
Thus, this relation between $\Phi-\Psi$ and $\delta$ is the same as that of the
general relativity. 
The signal of the weak lensing is determined by $\Phi-\Psi$ 
along the path of a light ray.
Therefore, we only consider the effect of the modified 
gravity on the matter density perturbations of 
Eq.~(\ref{evolutioneq}) for elaborating the weak lensing 
statistics.

%
In the present paper, the modified gravity of the $f(R)$ model
is supposed to be characterized by $n$ and $k_C$(or $\lambda$).
We perform the Fisher matrix analysis
with the $9$ parameters, $n$, $\lambda$(or $k_C$), 
$w_0,~w_a,~\Omega_0, ~\Omega_b, ~h, ~A,$ and $n_s$, where
$\Omega_b$ is the baryon density parameter, $n_s$ is the initial 
spectral index, $A$ is the amplitude of power spectrum. 
$w_0$ and $w_a$ characterize the background expansion history and the 
distance-redshift relation [see Eq.~(\ref{comogingdist})].
In the $f(R)$ model, the background expansion is consistently 
determined by the action (\ref{action}) once the form of $f(R)$
is specified. In the present paper, without specifying the 
explicit form of $f(R)$, we only adopted Eq.~(\ref{fR}) 
in an asymptotic region. 
And we assumed the $\Lambda$CDM model as the background 
expansion of the Universe in the previous section.
But we here consider possible uncertainties of the background 
expansion, by including the parameters $w_0$ and $w_a$.
However, as will be shown in the below, 
the inclusion of the parameters $w_0$ and $w_a$  
does not alter our result at the qualitative level.

\begin{figure}[t]
\begin{center}
\includegraphics[width=12cm,height=6cm,clip]{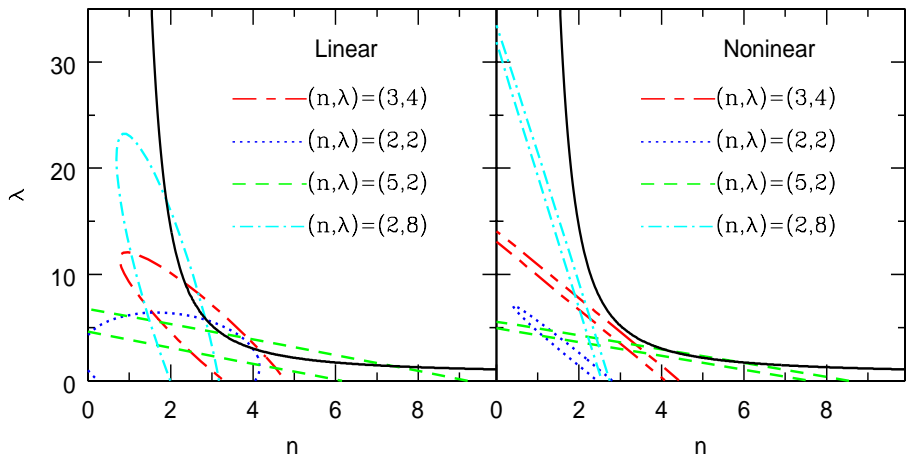}
\caption{ Left panel: The $1$-sigma contour in the $(n-\lambda)$ plane.
The linear modeling for $P_{\rm mass}(k,z)$ in the range of 
$10\leq l\leq 10^3$ is used. 
The target modes are $(n,\lambda)=(3,4), (2,2), (5,2)$, and $(2,8)$, 
respectively. 
The other target parameters are $w_0=-1,~w_a=0,~\Omega_0=0.28, 
~\Omega_b=0.044, ~h=0.7, \sigma_8=0.8$, and $n_s=0.95$.
The solid curve corresponds to $k_C=0.2h{\rm Mpc}^{-1}$.
Right panel: Same as the left panel, but with the nonlinear modeling 
for $P_{\rm mass}(k,z)$ of the range of $10\leq l\leq 3\times10^3$. 
}
\label{fig:n_lambda}
\end{center}
\end{figure}

\begin{figure}[htbp]
\begin{center}
\includegraphics[width=12cm,height=6cm,clip]{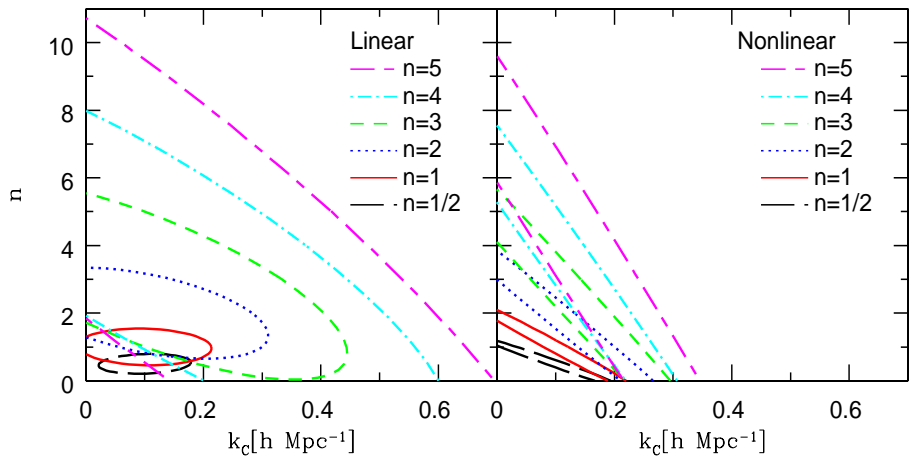}
\caption{ Left panel: The $1$-sigma contour in the $(k_C-n)$ plane.
The linear modeling for $P_{\rm mass}(k,z)$ in the 
range of $10\leq l\leq 10^3$ is used.
We assume the target modes $k_C=0.1h{\rm Mpc}^{-1}$ and 
$n=5,~4,~3,~2,~1$, and $1/2$, respectively, 
from the larger circle to smaller one.
The other parameters are the same as those of Fig.~\ref{fig:n_lambda}.
Right panel: Same as the left panel, but with the nonlinear modeling for 
$P_{\rm mass}(k,z)$ of the range of $10\leq l\ \leq 3\times10^3$. 
}
\label{fig:n_kC}
\end{center}
\end{figure}

\begin{figure}[htbp]
\begin{center}
\includegraphics[width=12cm,height=12cm,clip]{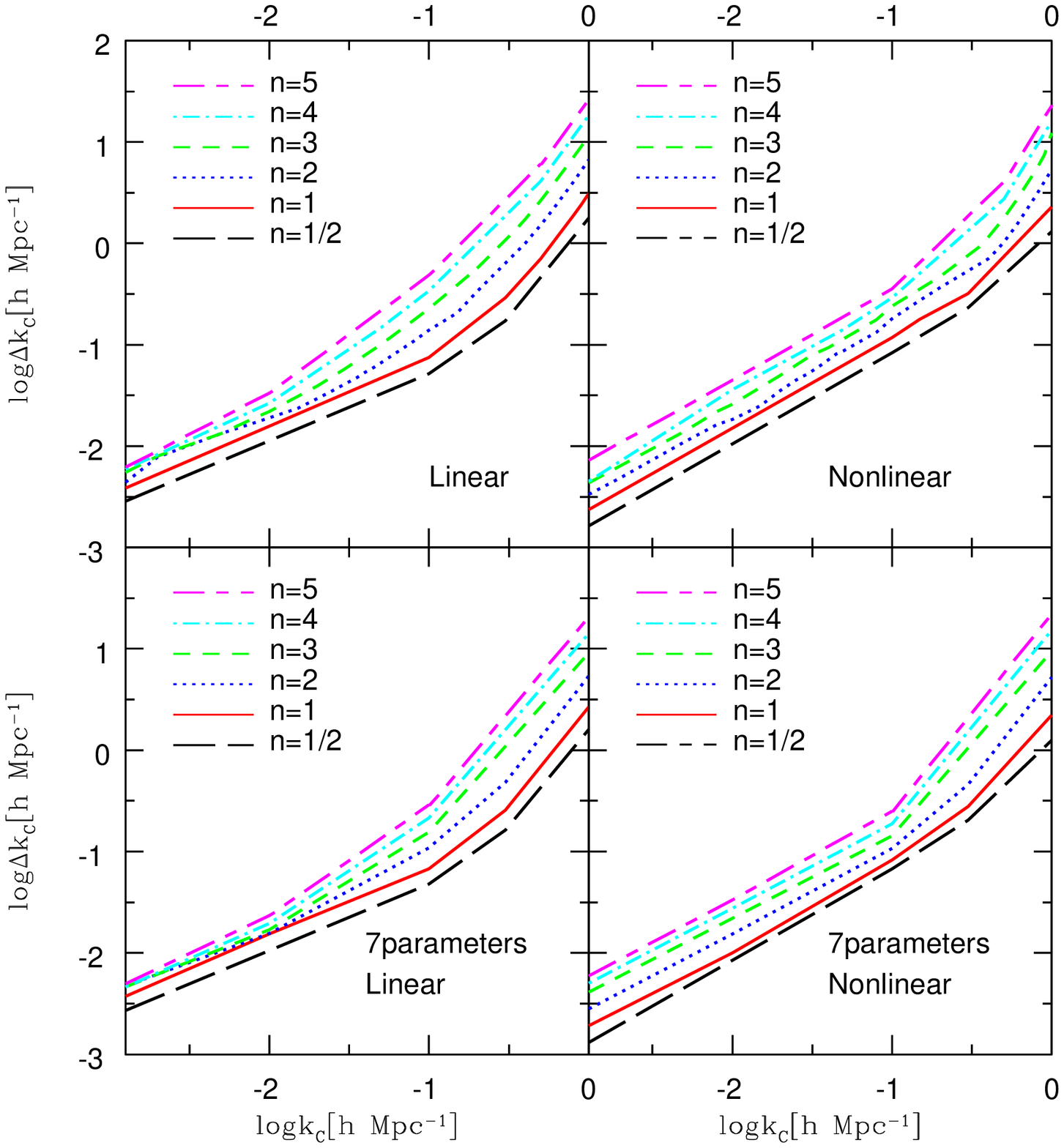}
\caption{The 1-sigma error on $k_C$ as a function of the target 
value of $k_C$, where the other parameters are marginalized over. 
The left (right) panels use the linear (nonlinear) modeling for 
$P_{\rm mass}(k,z)$ of the range of $10\leq l\leq 10^3$ 
($10\leq l\ \leq 3\times10^3$). In each panel, 
the curves assume the target parameter $n=5,~4,~3,~2,~1$, 
and $1/2$, from the top to the bottom, respectively. The other target 
parameters are the same as those of Fig.~\ref{fig:n_lambda}.
The upper panels are the result of the $9$ parameters, 
$k_C$, $n$, $w_0,~w_a,~\Omega_0, ~\Omega_b, ~h, ~A,$ and $n_s$.
The lower panels are the result of the $7$ parameters, 
$k_C$, $n$, $\Omega_0, ~\Omega_b, ~h, ~A,$ and $n_s$,
where the background expansion is fixed as that of the 
$\Lambda$CDM model.
}
\label{fig:Del_kC}
\end{center}
\end{figure}

In the Fisher matrix analysis, we assume the galaxy sample of a survey 
with the number density $N_g=35$ per arcmin.${}^2$, 
the mean redshift $z_m=0.9$, 
and the total survey area, $\Delta A=2\times 10^4$ square degrees. 
We also assumed the tomography with $4$ redshift bins (see also 
Appendix). 
Figure \ref{fig:n_lambda} is the result of the
Fisher matrix analysis of the $9$ parameters, $n$, $\lambda$,
$w_0,~w_a,~\Omega_0, ~\Omega_b, ~h, ~A,$ and $n_s$.
Figure \ref{fig:n_lambda} plots the $1$-sigma contour in the $n-\lambda$
plane, which is obtained by marginalizing the Fisher matrix over the other 
7 parameters.
The target values of $n$ and $\lambda$ are shown in the panels. 
The other target parameters are $w_0=-1,~w_a=0,~\Omega_0=0.28, 
~\Omega_b=0.044,~h=0.7$, $n_s=0.95$, 
and $A$ which is set so that $\sigma_8=0.8$. 
We take into account the Planck prior constraint of the expected errors
$\Delta w_0=0.6,~\Delta w_a=1.9,~\Delta\Omega_0=0.01, ~\Delta\Omega_b=0.0014,
 ~\Delta h=0.01, ~\Delta \sigma_8=0.1$ and $\Delta n_s=0.014$ 
\cite{Kitching08}.
The left panel shows the result using the linear theory for the
matter power spectrum of the range, $10\leq l \leq 10^3$. 
The right panel is the result with the nonlinear matter power spectrum
of the range, $10\leq l \leq 3\times 10^3$. 
In this figure, the solid curve corresponds to $k_C=0.2~h{\rm Mpc}^{-1}$, 
which was defined as the boundary between the general-relativity 
regime and the dispersion regime in the reference \cite{Tsujikawa09}. 

Figure \ref{fig:n_kC} is similar to Fig.~\ref{fig:n_lambda}, 
but the $1$-sigma contour in the $k_C-n$ plane. 
We assumed the target modes $k_C=0.1h{\rm Mpc}^{-1}$ and 
$n=5,~4,~3,~2,~1$, and $1/2$, from the larger circle to smaller one, respectively.
The other parameters are the same as those of Fig.~\ref{fig:n_lambda}.
Figure \ref{fig:Del_kC} shows the $1$-sigma error on $k_C$ 
as a function of the target value of $k_C$, where the 
other parameters are marginalized over. 
The left panels are the linear theory, while the right panels are the
nonlinear model. The upper panels are the result of the 
Fisher matrix of the $9$ parameters $n$, $k_C$,
$w_0,~w_a,~\Omega_0, ~\Omega_b, ~h, ~A,$ and $n_s$.
Then, the $1$-sigma error $\Delta k_C$ is evaluated 
by marginalizing the Fisher matrix over the 8 parameters 
$n,~w_0,~w_a,~\Omega_0, ~\Omega_b, ~h, ~A,$ and $n_s$.
The error of $k_C$ is the same order 
of $k_C$ for the cases $n=1/2$ and $1$, 
but the error becomes larger as $n$ becomes larger. 
The lower panels are the result of the 
Fisher matrix of the $7$ parameters, $k_C$, $n$, $\Omega_0,
~\Omega_b, ~h, ~A,$ and $n_s$, with fixing the background 
expansion to be that of the $\Lambda$CDM model.
Thus, the inclusion of the parameters $w_0$ and $w_a$ does not
alter the result qualitatively. 

Figures \ref{fig:n_lambda}-\ref{fig:Del_kC} show that the difference 
between the linear modeling and the nonlinear modeling is not very significant.
We adopted the Peacock and Dodds formula \cite{PeacockDodds96} 
for the nonlinear modeling of the matter power spectrum, while
the formula by Smith et al.~\cite{Smith03} has been used frequently 
\cite{Oyaizu08a,Oyaizu08b,Schmidt09a,Koyama09,Beynon09}. 
However, the choice of the nonlinear formula doesn't 
alter our conclusion qualitatively.
We have not taken the nonlinear effect from the 
Chameleon mechanism into account. The nonlinear modeling for the $f(R)$ model 
has not been studied well for the general case of $n$. 
The effect of the nonlinear modeling might need further investigations.

\section{summary and conclusions}

In the present paper, we have investigated the linear growth rate of 
cosmological matter density perturbations in the viable $f(R)$ model 
both numerically and analytically. 
We found that the growth rate in the scalar-tensor regime can be 
characterized by a simple analytic formula (\ref{STregime}). 
This is useful to understand the characteristic behavior of 
the growth index in the scalar-tensor regime. 
We also investigate a prospect of constraining the Compton
wavelength scale of the $f(R)$ model with a future weak lensing 
survey. This result shows that a constraint on $k_C$ of the same
order of $k_C$ will be obtained for the model $n=1$ and $n=1/2$, 
though the constraint is weaker as $n$ is larger. 
For $k_C\simgt 1h{\rm Mpc}^{-1}$, the constraint is very weak. 
This is because the weak lensing statistics is not very 
sensitive to the density perturbations on the smaller scales. 

\begin{figure}[htbp]
\begin{center}
\includegraphics[width=7cm,height=7cm,clip]{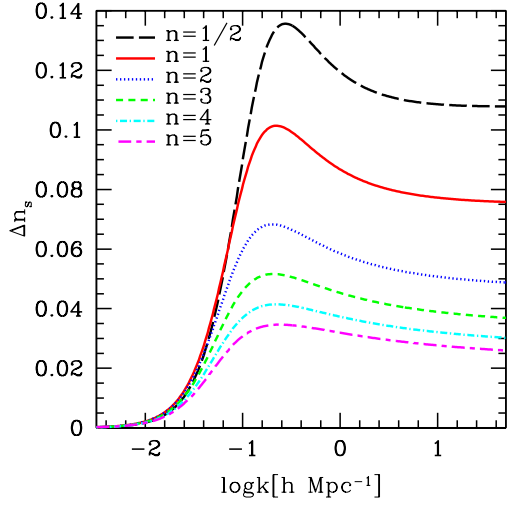}
\caption{{The additional spectral index $\Delta n_s$ as a 
function of wavenumber $k(h{\rm Mpc}^{-1})$. Here we adopted
$k_C=0.1h{\rm Mpc}^{-1}$. The curves assume $n=1/2,1,~2,~3,~4$, and $5$,
from the top to the bottom, respectively.}}
\label{fig:ns}
\end{center}
\end{figure}

Finally we mention about the effect of the late-time evolution 
of matter density perturbations in the $f(R)$ model on the
spectral index. This effect causes the additional spectral index, 
which is evaluated by $\Delta n_s={d\ln D_1^2(a,k)/d\ln k}$. 
The analytic formula of the additional spectral index is 
given by Starobinsky \cite{Starobinsky07} (see also 
\cite{Tsujikawa08,Motohashi}), $\Delta n_s={(\sqrt{33}-5)/(6n+4)}$, 
which yields $\Delta n_s=0.11,~0.074,~0.047,~0.034,~0.027$, and $0.022$, 
for $n=1/2,~1,~2,~3,~4$, and $5$, respectively. Figure \ref{fig:ns}
plots our numerical result of $\Delta n_s$ as a function of $k$ 
assuming $k_C=0.1h{\rm Mpc}^{-1}$. The numerical result
approaches the analytic result at $k\gg k_C$, but one can see 
the bump around the wavenumber $k_C$, depending on $n$. 
Possibility of detecting of the spectral shape is interesting, 
but is out of the scope of this paper. 

\acknowledgements
We thank Tsutomu Kobayashi, S. Tsujikawa, H. Motohashi, and 
J. Yokoyama for useful discussions.
We also thank R. Kimura for useful comments. 
This work is supported by Japan Society for Promotion
of Science (JSPS) Grants-in-Aid
for Scientific Research (No.~21540270, No.~21244033).
This work is also supported by JSPS 
Core-to-Core Program ``International Research 
Network for Dark Energy''.
T.N. acknowledges support by a research assistant program
of Hiroshima University. 
We utilized the MATHEMATICA~6.0 in parts of our investigations.

\appendix
\section{Modeling of weak lensing survey power spectrum}
We briefly review the Fisher matrix analysis for
a weak lensing survey. 
The analysis in the present paper is almost the same
as that of Ref.~\cite{Yamamoto07}, but the difference is
the modeling for the evolution of the matter density perturbations. 
%

As is described in Sec.~I\hspace{-.1em}V, the signal of the weak lensing is 
determined by $\Phi-\Psi$ along the path of a light ray. 
Assuming the weak lensing tomography method \cite{WHU},
the cosmic shear power spectrum for the $i$-th and $j$-th 
redshift bins is 
\begin{equation}
 P_{(ij)}(l)=\int_0^\infty d\chi\overline{W}_i(z(\chi))\overline{W}_j(z(\chi))
\left({l\over\chi}\right)^4{1\over 4}
P_{\Phi-\Psi}\left(k\rightarrow\frac{l}{\chi},z(\chi)\right),
\label{Powerlens}
\end{equation}
where $P_{\Phi-\Psi}(k,z)$ is the power spectrum of $\Phi-\Psi$, 
$\chi$ is the comoving distance, 
$\overline{W}_i(z)$ is the weight factor of the $i$-th redshift bin,
\begin{equation}
 \overline{W}_i(z)=\frac{1}{\overline{N}_i}\int_{{\rm max}(z_i,z)}^{z_i+1}dz'\frac{dN(z')}{dz'}\left(1-\frac{\chi(z)}{\chi(z')}\right),
\end{equation}
where $dN/dz(z)$ denotes the differential number count of galaxies 
with respect to redshift per unit solid angle, and 
$\overline{N}_i=\int_{z_i}^{z_i+1}dz'(dN(z')/dz')$ is the total number of galaxies
in the $i$-th redshift bin. From Eq.~(\ref{PPd}), Eq.~(\ref{Powerlens})
is written as
\begin{equation}
 P_{(ij)}(l)=\int_0^\infty d\chi\overline{W}_i(z(\chi))\overline{W}_j(z(\chi))
\left(\frac{3H_0^2\Omega_0}{2a}\right)^2
P_{\rm mass}\left(k\rightarrow\frac{l}{\chi},z(\chi)\right),
\label{Plwer}
\end{equation}
where $P_{\rm mass}(k,z)$ is the matter power spectrum, 
for which we adopted the Peacock and Dodds formula \cite{PeacockDodds96}
for the nonlinear modeling. This expression (\ref{Plwer}) 
is familiar
as the weak shear power spectrum, but the modification of the gravity 
is involved in the evolution of 
the matter power spectrum $P_{\rm mass}(k,z)$.

In the present paper, we adopt the comoving distance 
\begin{equation}
 \chi(z)=\int_0^z{dz'\over H(z')}
=\int_0^z{dz'\over H_0\sqrt{\Omega_0(1+z')^3+(1-\Omega_0)(1+z')^{3(1+w_0+w_a)}e^{-3w_az'/(1+z')}}},
\label{comogingdist}
\end{equation}
which includes $w_0$ and $w_a$, the parameters
of the equation of state of the dark energy 
$w(z)=w_0+w_a(1-a)$. 
As mentioned in Sec.~I\hspace{-.1em}V, the background expansion of
the $f(R)$ model is specified once the form of $f(R)$
is given. The $f(R)$ model, in the present paper, only 
assumes the form in an asymptotic region. Taking possible
uncertainties of the background expansion, we include 
$w_0$ and $w_a$ in the Fisher matrix analysis. 
However, this inclusion does not alter our result. 

The fisher matrix is estimated as
\begin{equation}
 F_{\alpha\beta}=\sum_l\sum_{(ij)(mn)}\frac{\partial P_{(ij)}}
{\partial \theta_\alpha}{\rm Cov}_{(ij)(mn)}^{-1}\frac{\partial P_{(mn)}}
{\partial \theta_\beta},
\end{equation}
where $\theta^\alpha$ is a parameter of the theoretical modeling, 
and the covariance matrix is 
\begin{equation}
 {\rm Cov}_{(ij)(mn)}(l) =  \frac{1}{(2l+1)\Delta lf_{\rm sky}}\left[P_{(im)}^{\rm obs}(l)P_{(jn)}^{\rm obs}(l)+P_{(in)}^{\rm obs}(l)P_{(jm)}^{\rm obs}(l)\right],
\end{equation}
with $ P_{(ij)}^{\rm obs}(l)=P_{(ij)}(l)+\delta_{ij}{\sigma_\epsilon^2}
/{\overline{N}_i}$,
where $f_{\rm sky}$ is the fraction of the survey area, and $\sigma_\epsilon$
is the rms value of the intrinsic random ellipticity, which we take 
$0.22$. 
In the Fisher matrix analysis, we assume the sample of galaxies 
of imaging survey modeled as
\begin{equation}
{dN\over dz}={N_g\beta\over z_0^{\alpha+1} \Gamma((\alpha+1)/\beta)}
z^{\alpha}\exp\left[-\left({z\over z_0}\right)^\beta\right],
\end{equation}
with $\alpha=0.5$, $\beta=3$, $N_g=35$ per arcmin${}^2$, and $z_0$ is 
given by the relation, 
$z_0=z_m\Gamma((\alpha+1)/\beta)/\Gamma((\alpha+2)/\beta)$ 
so that the mean redshift is $z_m=0.9$. 
We assume the survey area, $\Delta A=2\times 10^4$ square degrees, 
and the tomography with $4$ redshift bins.


\begin{thebibliography}{999}
\bibitem{footnote1} {After we have completed this manuscript, 
we noticed the paper by Brax et al.\cite{Brax09}, in which 
the similar formula for the growth rate in the scalar-tensor regime 
is presented.}
\end{thebibliography}
\end{document}